\documentclass[preprintnumbers,amsmath,amssymb,twocolumn,nofootinbib,aps,prl]{revtex4}
\usepackage{amsfonts} \usepackage{amssymb} \usepackage{latexsym}
\usepackage{graphicx}
\usepackage[utf8x]{inputenc} \usepackage{longtable}
\usepackage{dcolumn} \usepackage{amsmath} \usepackage{bm}
\usepackage{young}

\usepackage[active]{srcltx} 

\newcommand{\al}{\alpha}

\begin{document}

\title{Stable He$^-$ can exist in a strong magnetic field}

\author{A.V.~Turbiner} \email{turbiner@nucleares.unam.mx}

\author{J. C. Lopez Vieyra} \email{vieyra@nucleares.unam.mx}
\affiliation{Instituto de Ciencias Nucleares, Universidad Nacional
  Aut\'onoma de M\'exico, Apartado Postal 70-543, 04510 M\'exico,
  D.F., Mexico}

\begin{abstract}
The existence of bound states of the system $(\al,e,e,e)$ in a magnetic field $B$ is studied
using the variational method. It is shown that for $B \gtrsim 0.13\,{\rm a.u.}$ this system gets bound with total energy below the one of the $(\al,e,e)$ system.
It manifests the existence of the stable He$^-$ atomic ion. Its ground state is a spin-doublet $^2(-1)^{+}$ at $0.74\, {\rm a.u.} \gtrsim B \gtrsim 0.13\, {\rm a.u.}$ and it becomes
a spin-quartet $^4(-3)^{+}$ for larger magnetic fields. For $0.8\, {\rm a.u.} \gtrsim B \gtrsim 0.7\, {\rm a.u.}$ the He$^-$ ion has two (stable) bound states $^2(-1)^{+}$ and $^4(-3)^{+}$.
\end{abstract}

\maketitle


Astrophysical objects like magnetic white dwarfs may have surface
magnetic fields of $10^8 - 10^{10}$\,Gauss ($\sim 0.1 - 10$\,a.u.) whilst
neutron stars typically reach $10^{12} - 10^{13}$\, Gauss ($\sim 10^3 -
10^4$\,a.u.) or even $10^{15}$\, Gauss ($\sim 10^6$\,a.u.) in the so
called magnetars.  In the presence of such strong magnetic fields the
chemical properties of atoms and molecules change dramatically.  In particular,
it makes possible the formation of unusual chemical compounds
like the H$_3^{++}$ ion at $B \gtrsim 10^{10}$\,G \cite{Turbiner:1999}
(for a review see \cite{Turbiner:2006} about one-electron molecular systems,
and \cite{Turbiner:2010} about two-electron atomic-molecular systems,
and references therein).
A separate question of interest concerns the existence of negative atomic ions
in a magnetic field. An immediate observation is that the induced quadrupole moment
- charge interaction of the atomic core with an electron is repulsive: it can
influence binding.  Thus, it was quite a striking theoretical result that the
simplest negative atomic ion  ${\rm H}^{-}$, which possesses the single bound
state~\cite{Hill:1977}, develops an infinite number of bound states
in the presence of a magnetic field~\cite{Avron:1977}.  A similar
situation may occur for the case of the negative atomic ion
$\rm{He}^{-}$ which does not seem to have a stable bound state in the field-free
case~\cite{Andersen:2004, King:2008}, but can become bound in a magnetic
field.

The goal of this paper is to explore the possibility of having
stable bound states of the 1-center Coulomb system $(\al,e,e,e)$ in a
magnetic field checking the existence of the negative ion He$^-$.
Our main motivation to study the negative ion He$^-$ in a magnetic field comes
from the recently observed spectra of white dwarfs which indicate the presence of the
atomic Helium on the surface of some of these astrophysical magnetized
objects, see e.g.~\cite{Jordan:1998}. Therefore, the existence of ${\rm He}^{-}$
ions can be of relevance to interpret the observed absorption features in the spectra.
In this paper atomic units ($\hbar=e=m_e=1$) are used throughout, and the magnetic field
is measured in units of $B_0 = 2.35\times 10^9$\, G.

The non-relativistic Hamiltonian for a three-electron, one-center system in a
magnetic field (directed along the $z$-axis and taken in the symmetric gauge) with an
infinitely massive nucleus is
\begin{eqnarray}
\label{H}
  {\cal H}\ & = & - \sum_{k=1}^3 \left( \frac{1}{2} {\nabla}_{k}^2\ +
  \  \frac{Z}{r_{k}} \right) \ + \ \sum_{{k=1}}^3 \sum_{{j>k}}^3
  \ \frac{1}{r_{kj}}\  \nonumber \\ && +\frac{B^2}{8} \sum_{k=1}^{3}
  \ {\rho_k}^2  + \frac{B}{2} (\hat L_z +2\hat S_z ) \, ,
\end{eqnarray}
where ${\nabla}_{k}$ is the 3-vector momentum of the $k$th electron,
$r_{k}$ is the distance between the $k$th electron and the nucleus,
$\rho_k$ is the distance of the $k$th electron to the $z$-axis, and
$r_{kj}$ \,$(k,j=1,2,3)$ are the inter-electron distances. $\hat L_z$
and $\hat S_z$ are the $z$-components of the total angular momentum and
total spin operators, respectively. Both $\hat L_z$ and $\hat S_z$ are
integrals of motion and can be replaced in (\ref{H}) by their eigenvalues $M$ and $S_z$
respectively.  $Z$ is the nuclear charge (for He$^-$ $Z$=2).  The total
spin ${\hat S}$ and $z$-parity $\hat{\Pi}_z$ are also conserved
quantities.  The spectroscopic notation $\nu{}^{2S+1}M^{\Pi_z}$ is
used to mark the states, where $\Pi_z$ denotes  the $z$~parity eigenvalue
$(\pm)$, and  the quantum number $\nu$ labels the degree of excitation.
For states with the same symmetry, for the lowest energy states at $\nu=1$
the notation is ${}^{2S+1}M^{\Pi_z}$. We always consider states with $\nu=1$
and $S_z=-S$ assuming they correspond to the lowest total energy states of
a given symmetry in a magnetic field.

The variational method is used to explore the problem. The recipe
for choosing the trial function is based on arguments of physical
relevance: the trial function should support the symmetries of
the system, has to reproduce the Coulomb singularities of the potential correctly
and the asymptotic behavior at large distances (see,
e.g. \cite{turbinervar, turbinervar1, Turbiner:2006}). It implies
that electron-electron interaction plays an
important role, thus, the correlation should be introduced into trial
functions in exponential form $\sim \exp(\al \, r_{ij})$, where
$\al$ is a variational parameter.

Following the above, a trial function for the spin 1/2 lowest energy state
is chosen in the form
\begin{equation}
\label{GStrialfunct}
 \psi(\vec{r}_1,\vec{r}_2,\vec{r}_3) = {\cal A} \left[ \,
   \phi(\vec{r}_1,\vec{r}_2,\vec{r}_3)\chi \, \right]\,,
\end{equation}
where $\chi$ is the spin eigenfunction, ${\cal A}$ is
the three-particle antisymmetrizer
\begin{equation}\label{Asym}
 {\cal A} = 1 - P_{12} - P_{13} - P_{23} + P_{231}  + P_{312}\,,
\end{equation}
and  $\phi(\vec{r}_1,\vec{r}_2,\vec{r}_3)$ is the explicitly
correlated orbital  function
\begin{eqnarray}
\label{Phitrialfunct}
\phi(\vec{r}_1,\vec{r}_2,\vec{r}_3)  &=& \left( \prod_{k=1}^{3}
\rho_k^{|M_k|} e^{ i M_k \phi_k} \, e^{-\al_k r_k - \frac{B}{4}
  \beta_k \rho_k^2} \right) \times \nonumber \\[5pt] &&
e^{\scriptstyle \al_{12} r_{12} + \al_{13} r_{13} + \al_{23}
  r_{23}} \,,
\label{psi32}
\end{eqnarray}
where $M_k$ is the magnetic quantum number of the $k$-th electron, and
$\al_k$, $\beta_k$ and $\al_{kj}$ are non-linear variational
parameters. In total, the trial function (\ref{GStrialfunct}) contains
9 variational parameters.  The function (\ref{GStrialfunct}) is a
properly anti-symmetrized product of $1s$ Slater type orbitals, the lowest
Landau orbitals and the exponential correlation factors $\sim\exp{(\al\, r_{kj})}$.
We expect the ground state to be realized by different states depending on the domain
of magnetic fields: guided by an analogy with the case of the lithium atom
in a magnetic field (for a discussion see \cite{SchmelcherLi:1998}), we assume
the spin 1/2 states $^2(0)^{+}$, $^2(-1)^{+}$ to correspond to the ground state for
small and intermediate magnetic fields, respectively, while the spin 3/2 state
$^4(-3)^{+}$ is the ground state for the large magnetic fields.

For the states ${}^2(0)^{+}$, ${}^2(-1)^{+}$ of the total spin $S=1/2$
we have two linearly independent spin functions of mixed symmetry
\begin{equation}
\label{chi1}
 \chi_1 =  \frac{1}{\sqrt{2}} [ {\boldsymbol\alpha}(1)
   {\boldsymbol\beta}(2)    - {\boldsymbol\beta}(1)
   {\boldsymbol\alpha}(2) ]{\boldsymbol\alpha}(3)
\end{equation}
and
\begin{equation}
\label{chi2}
 \chi_2 =  \frac{1}{\sqrt{6}} [  2{\boldsymbol\alpha}(1)
   {\boldsymbol\alpha}(2)  {\boldsymbol\beta}(3)   -
   {\boldsymbol\beta}(1)  {\boldsymbol\alpha}(2)
   {\boldsymbol\alpha}(3)  - {\boldsymbol\alpha}(1)
   {\boldsymbol\beta}(2) {\boldsymbol\alpha}(3) ]\,,
\end{equation}
where ${\boldsymbol\alpha}(i)$, ${\boldsymbol\beta}(i)$ are spin up,
spin down eigenfunctions of the $i$-th electron. The spin function $\chi$ in
(\ref{GStrialfunct}) is chosen as
\[
   \chi\ =\ \chi_1 + c \chi_2 \ ,
\]
(for discussions see \cite{TGH2009} and \cite{Drake:2012}), where $c$ is variational parameter.
For entire range of studied magnetic fields, $c$ is different but close to zero.
For the spin $S=3/2$  state ${}^4(-3)^{+}$ with $M_1+M_2+M_3=-3$,
the spin part corresponds to the totally symmetric spin function $\chi={\boldsymbol\beta}(1){\boldsymbol\beta}(2){\boldsymbol\beta}(3)$,
and the orbital part $\phi(\vec{r}_1,\vec{r}_2,\vec{r}_3) $ is anti-symmetrized
by applying the operator ${\cal A}$  (Eq. (\ref{Asym})).

The variational energy has quite complicated profile in the parameter space: use
of standard minimization strategies did not allow us to find a minimum
reasonably fast. As a result most of the minimization was performed manually
using the procedure SCAN from the minimization package MINUIT from CERN-LIB.
Nine-dimensional integrals which appear in the functional of energy are calculated
numerically using a "state-of-the-art" dynamical partitioning procedure based on
division of the integration domain following the profile of the integrand,
separating domains with large gradients.
Each subdomain was integrated separately in parallel and with
controlled absolute/relative accuracy (for details, see
e.g. \cite{Turbiner:2006}). Numerical integration of every subdomain
is done with a relative accuracy \hbox{of $\sim 10^{-2} - 10^{-4}$}
using an adaptive routine \cite{GenzMalik}.  Parallelization is
implemented using the MPI library MPICH.  Computations are performed
on a Linux cluster with 96 Xeon processors at 2.67\,GHz each, and
12Gb RAM.

The existence of a chemical compound is established when
the system possesses at least one stable bound state. If such a bound
state exists, it is characterized by a positive ionization energy,
{\it i.e.}  the minimal amount of energy which is necessary to add to
the system to separate it into two or more sub-systems. In
particular, the one-particle ionization energy is defined as the
energy needed to move an electron to infinity. A bound state of He$^-$ is
characterized by definite values of the total magnetic
quantum number and $z$~projection of the total spin $(M,S_{z})$. Then,
such a state is stable if its total energy is smaller than the sum
of energies of two sub-systems (He-atom + $e$) {\it i.e.} if
\begin{equation}
\label{estabilitycondition}
  E_T^{\mbox{\scriptsize He}^-}(M,S_z)  <    E_T^{\mbox{\scriptsize
      He}}(M^\prime,S_{z}^\prime)  + E_T^{\mbox{\scriptsize
      e}^-}(M_{e^-},S_{z_{e^-}})\,,
\end{equation}
where $E_T^{\mbox{\scriptsize He}}(M^\prime,S_{z}^\prime)$ and
$E_T^{\mbox{\scriptsize e}^-}(M_{e^-},S_{z_{e^-}})$ are the total
energies of the He-atom and the electron, respectively (see
Ref.~\cite{SchmelcherLi:2004}). The condition (\ref{estabilitycondition})
must be valid for all possible decay channels satisfying the conservation
of the quantum numbers
\begin{equation}
\label{MSzconservation}
M     =   M^\prime + M_{e^-} \,, \quad S_z  =   S_{z}^\prime  +
S_{z_{e^-}}  \,,
\end{equation}
which are valid in the non-relativistic approximation.
For an electron in a magnetic field, the total energy of the Landau
levels is given by
\begin{equation}
 E_T^{\mbox{\scriptsize e}^-}(M_{e^-},S_{z_{e^-}}) = ( M_{e^-} +
 |M_{e^-}| + 2 S_{z_{e^-}} + 1 )\ \frac{B}{2}\,,
\end{equation}
and for non-positive values of the magnetic quantum number
$M_{e^-}$, the Landau levels are infinitely degenerate:
\begin{equation}
 E_T^{\mbox{\scriptsize e}^-}(M_{e^-}\leq 0,S_{z_{e^-}}) = ( 2
 S_{z_{e^-}}  + \, 1 )\ \frac{B}{2}\ .
\end{equation}
Hence, for an electron with $z$-spin projection antiparallel to the
magnetic field and zero (or negative) magnetic quantum number:
$E_T^{\mbox{\scriptsize e}^-}(M_{e^-}\leq 0,S_{z_{e^-}}=-1/2) =0$,
whereas an electron with $z$-spin projection parallel to the magnetic
field and zero or negative magnetic quantum number has
$E_T^{\mbox{\scriptsize e}^-}(M_{e^-}<0,S_{z_{e^-}}=+1/2) = B$.

\begin{figure}[tb]
\begin{center}
   \includegraphics*[width=3.4in,angle=0]{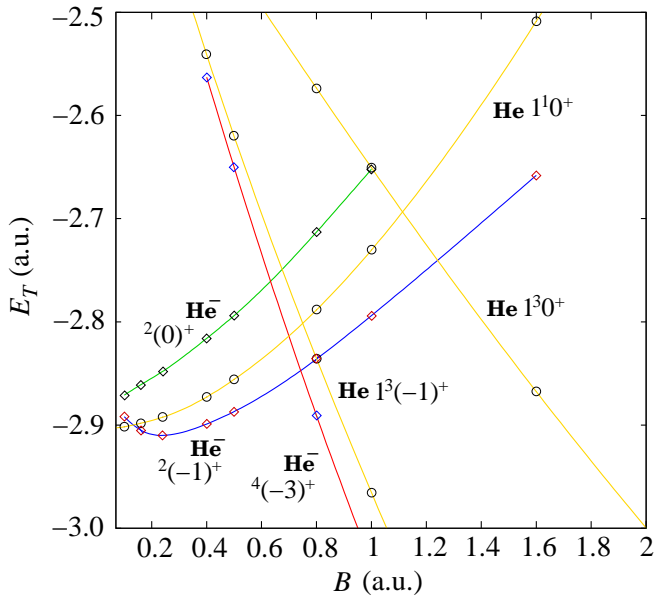}
    \caption{\label{Figure1} Total energies (in a.u.) for the states
      $^2(0)^{+}$, $^2(-1)^{+}$ and  $^4(-3)^{+}$ of the negative
      ion He$^-$  (open diamonds) in comparison to the energies  of
      the lowest He-atom states $1{}^1 0^+$, $1{}^3 0^+$, $1{}^3 (-1)^+$
      (open circles) in a  magnetic field $B$.}
    \end{center}
\end{figure}



{\bf State ${}^2(0)^{+}$}\ .
The state ${}^2(0)^{+}$ of the system $(\al,e,e,e)$ in a magnetic
field of strength $B$  is described by the trial function
(\ref{GStrialfunct}) with $M_1=M_2=M_3=0$ (see
(\ref{Phitrialfunct})).  Since this state ${}^2(0)^{+}$
is the ground state for the lithium atom for weak magnetic
fields~\cite{SchmelcherLi:1998},  it is natural to assume that
the system  $(\al,e,e,e)$ can also develop a stable
ground state with this symmetry in a magnetic field. However,
our results show that the total energy of this state always lies well above
the ${\rm He}$ ground state $ 1{}^10^+$ in the whole domain of magnetic fields
studied (see Table~\ref{Table1} and Fig.~\ref{Figure1}).
Thus, this state is metastable being unstable towards decay
${\rm He}^- ({}^2(0)^{+}) \to$ He $(1{}^10^+) +\, e$.
The total energy of this metastable state grows monotonically with an increase
of the magnetic field strength (see Table~\ref{Table1} and Figure~\ref{Figure1}).

{\bf State   ${}^2(-1)^{+}$}\ .
The state ${}^2(-1)^{+}$ of the system $(\al,e,e,e)$ in a magnetic
field is described by the trial function (\ref{GStrialfunct}) with
$M_1=M_2=0, M_3=-1$ (see (\ref{Phitrialfunct})). This state becomes
the lowest energy (ground) state of the lithium atom for intermediate magnetic
fields~(see e.g. \cite{SchmelcherLi:1998}). The state  ${}^2(-1)^{+}$ for
He$^-$ gets bound for all magnetic fields studied. Its variational energies
are shown at Table~\ref{Table1} for 1.6~a.u.~$> B > 0.1$\,a.u. These energies
are always below the total energies of the state ${}^2(0)^{+}$. Qualitatively,
the total energy of the state ${}^2(-1)^{+}$ displays a minimum  at $B\simeq
0.25\,{\rm a.u.}$ and then grows monotonically with a magnetic field increase
(see Fig.~\ref{Figure1}). For magnetic fields $B \gtrsim 0.13\, {\rm a.u.} $
this state turns out to be stable towards decay
${\rm He}^- ({}^2(-1)^{+}) \to {\rm He} (1{}^10^{+})  + \, e$,
since its total energy lies below the energy of the He ground state $1{}^10^+$
for all magnetic fields (see Fig.~\ref{Figure1}). However, for magnetic fields
$B\gtrsim 0.8$\,a.u. the state  ${}^2(-1)^{+}$ of the system $(\al,e,e,e)$ becomes metastable: $(\al,e,e,e)$ decays to $(\al,e,e) + e$.

Hence, the state ${}^2(-1)^{+}$ realizes the stable bound state of the system $(\al,e,e,e)$ for magnetic fields $0.8\, {\rm a.u.} \gtrsim B \gtrsim 0.13\, {\rm a.u.}$. Eventually, it becomes
the ground state of the He$^-$-ion for $0.74\, {\rm a.u.} \gtrsim B \gtrsim 0.13\, {\rm a.u.}$ and
the first (stable) excited state for $0.8\, {\rm a.u.} \gtrsim B \gtrsim 0.74\, {\rm a.u.}$
(see below).



{\bf State   ${}^4(-3)^{+}$}\ .
The spin 3/2 state ${}^4(-3)^{+}$ of the system $(\al,e,e,e)$ in a magnetic field is
described by the trial function (\ref{GStrialfunct}) with $M_1=0, M_2=-1, M_3=-2$ (see
(\ref{Phitrialfunct})).
Due to the spin Zeeman contribution, the energy of this (spin $S=3/2$)
state decreases rapidly and monotonically with the magnetic field increase
(see Fig.~\ref{Figure1} and Table~\ref{Table1}). Based on pure energy considerations,
one can see that the system $(\al,e,e,e)$ in the state ${}^4(-3)^{+}$ gets stable for
$B\gtrsim 0.7$\,a.u.

At $B\simeq 0.7\, {\rm a.u.}$ the total energy of the state
${}^4(-3)^{+}$ of He$^-$ coincides with the total energy of the state $ 1{}^10^{+}$
of the He-atom. Hence, this state becomes the first excited state of the He$^-$ ion, while the ground state is ${}^2(-1)^{+}$.
The total energy of the state ${}^4(-3)^{+}$ continues to decrease with the magnetic
field increase. At $B \simeq 0.74\,{\rm  a.u.}$ it intersects with the total energy
of the state ${}^2(-1)^{+}$ and becomes the ground state of the He$^-$ ion for larger magnetic fields. In the domain $0.8\,{\rm  a.u.} \gtrsim  B \gtrsim 0.74\,{\rm  a.u.}$ the He$^-$ ion has two stable states: ${}^4(-3)^{+}$ as the ground state and ${}^2(-1)^{+}$ as the first excited state.
At $B \simeq 0.8\,{\rm  a.u.}$ the latter state gets metastable decaying to ${\rm He} (1{}^3 (-1)^{+}) + e$. Thus, for larger magnetic fields $B \gtrsim 0.8\,{\rm  a.u.}$ the He$^-$ ion has a single stable bound state ${}^4(-3)^{+}$.

\begin{table*}
\centering
\begin{tabular}{| l |c | c |c || l | l | l |}
\hline
&&&&&& \\[-7pt]
&He$^-$& He$^-$ & He$^-$& He  & He & He  \\ \hline
&&&&&& \\[-7pt]
$B$ a.u. &  $E(\,{}^2(0)^+)$ & $E(\,{}^2(-1)^+)$  & $E(\,{}^4 (-3)^+)$ &  $E(1\,{}^10^+)$  &   $E({1}^3\,{0}^+) $ &   $E(1\,{}^3 (-1)^+)$ \\
&$\scriptstyle M=0,\, S_z=-1/2$&$\scriptstyle M=-1\, S_z=-1/2$& $\scriptstyle M=-3,\, S_z=-3/2$
&$\scriptstyle M=0,\, S_z=0$ & $\scriptstyle M=0,\, S_z=-1$  & $\scriptstyle M=-1,\, S_z=-1$  \\
\hline\hline
0.1   & -2.871  &-2.892 &        & -2.901 740\footnotemark[4]   & -2.258 237\footnotemark[4] & -                           \\ \hline
0.16  & -2.861  &-2.905 &        & -2.898 290\footnotemark[1]   & -2.296 318\footnotemark[1] & -2.325 189\footnotemark[2]  \\ \hline
0.24  & -2.848  &-2.904 &        & -2.892 404\footnotemark[4]   & -2.339 571\footnotemark[4] & -2.402 393\footnotemark[2]  \\ \hline
0.40  & -2.816  &-2.899 & -2.563 & -2.872 874\footnotemark[4]   & -2.412 731\footnotemark[4] & -2.540 763\footnotemark[2]  \\ \hline
0.50  & -2.794  &-2.887 & -2.650 & -2.855 859\footnotemark[1]   & -2.454 347\footnotemark[1] & -2.620 021\footnotemark[2]  \\ \hline
0.8   & -2.713  &-2.836 & -2.891 & -2.788 425\footnotemark[4]   & -2.573 620\footnotemark[4] & -2.835 619\footnotemark[2]  \\ \hline
1.0   & -2.652  &-2.794 & -3.034 & -2.730 373\footnotemark[4]   & -2.650 658\footnotemark[4] & -2.965 504\footnotemark[2]  \\ \hline
1.6   &         &-2.658 & -3.394 & -2.508 81 \footnotemark[4]   & -2.867 620\footnotemark[1] & -3.308 774\footnotemark[2]  \\ \hline
2.0   &         &       & -3.606 & -2.330 65 \footnotemark[4]   & -2.999 708\footnotemark[1] & -3.508 911\footnotemark[2]  \\ \hline
5.0   &         &       & -4.764 & -0.575 5  \footnotemark[4]   & -3.768 199\footnotemark[1] & -4.625 491\footnotemark[2]  \\ \hline
10.0  &         &       & -5.999 &  3.064 582\footnotemark[1]   & -4.627 450\footnotemark[1] & -5.839 475\footnotemark[2]  \\ \hline
20.0  &         &       & -7.614 & 11.267 051\footnotemark[1]   & -5.772 448\footnotemark[1] & -7.440 556\footnotemark[2]  \\ \hline
50.0  &         &       & -10.46 & 38.076 320\footnotemark[1]   & -7.815 256\footnotemark[1] & -10.284 10\footnotemark[2]  \\ \hline
100.0 &         &       & -13.29 & 84.918 313\footnotemark[1]   & -9.843 074\footnotemark[1] & -13.104 78\footnotemark[2]  \\ \hline
\end{tabular}
\caption{\label{Table1}
Total energies in a.u. (Hartrees) for the states ${}^2(0)^+$, ${}^2(-1)^+$ and  ${}^4 (-3)^+$ for the system $(\al,e,e,e)$ obtained with trial function~(\ref{GStrialfunct}). For comparison, the total energies of the He-atom states  $1\,{}^10^+$, ${1}^3\,{0}^+ $ and $1\,{}^3 (-1)^+$ are included.
}
\footnotetext[1]{Ref. \cite{HeBecken:1999}, Becken: 1999}
\footnotetext[2]{Ref. \cite{HeBecken:2000}, Becken: 2000 }
\footnotetext[4]{Ref. \cite{Hesse:2004}, Hesse: 2004 }
\end{table*}


{\bf Lithium}\ .
In order to have an independent  estimate of the accuracy reached, we
have made some test calculations with the trial function
(\ref{GStrialfunct}) for the ${}^20^+$, ${}^2(-1)^+$ and for tightly bound
${}^4 (-3)^+$ states of the lithium atom in a magnetic field. Our results
are presented in the Table~\ref{Table2} where we include  previous
results~\cite{SchmelcherLi:2004} to compare with.

\begingroup
\squeezetable
 \begin{table}
\centering
\begin{tabular}{ |l|c|c|c|c|c|c| }
\hline
 $B$ a.u. &\multicolumn{2}{|c|}{$E(1\,{}^2 0^{+})$}  &  \multicolumn{2}{|c|}{$E(1\,{}^2 (-1)^{+})$} &  \multicolumn{2}{|c|}{$E(1\,{}^4 (-3)^+)$} \\ \hline
          & (\ref{GStrialfunct})  & \cite{SchmelcherLi:2004} &  (\ref{GStrialfunct})
          & \cite{SchmelcherLi:2004} & (\ref{GStrialfunct})  & \cite{SchmelcherLi:2004}\\
 \hline\hline

  0.0   & -7.455 \footnotemark[5] & -7.477766\footnotemark[6]  & -7.406  & -7.407126\footnotemark[6]           &            &  -5.142319\footnotemark[6]        \\ \hline
  1.0   &                         & -7.458550\footnotemark[6]  & -7.701  & -7.716679\footnotemark[6]           &  -6.567  &  -6.582361\footnotemark[6]        \\ \hline
  5.0   &                         & -6.136918\footnotemark[6]  &           & -7.002346\footnotemark[6]         &            &  -9.591769\footnotemark[6]   \\ \hline
\end{tabular}
\caption{\label{Table2} Total energies in a.u. for Li atom in a magnetic field for the states
${}^2 (-1)^+$,  ${}^2 (-1)^{+}$ and ${}^4 (-3)^+$ obtained with (2) and compared with
  \cite{SchmelcherLi:2004}.}
\footnotetext[5]{Ref. \cite{TGH2009}, Turbiner: 2009}
\footnotetext[6]{Ref. \cite{SchmelcherLi:2004}, Al-Hujaj: 2004}
\end{table}
\endgroup

\subsection{Conclusions}
We have shown that the system $(\al,e,e,e)$ in a magnetic field has
at least one stable bound state for magnetic fields $B \gtrsim 0.13\,{\rm a.u.}$ This
manifests the existence of the stable He$^-$ atomic ion.
For values of the magnetic field in $0.80 \gtrsim B \gtrsim 0.70\,{\rm a.u.}$
the system displays two stable bound states with the ground  state being realized
at first by the state ${}^2 (-1)^+$ for magnetic fields up to $B \simeq 0.74\,{\rm a.u.}$,
followed by the state ${}^4 (-3)^+$ as the stable ground state for $B \gtrsim  0.74\,{\rm a.u.}$, while the state  ${}^2 (-1)^+$ becomes the excited state.
For magnetic fields $B\gtrsim~0.80$\,~a.u. the negative ion ${\rm He}^-$ has a single stable bound state. All this shows that the closed shell argument does not work in a magnetic field.

It is worth noting that the energy of bound-free transitions grows very slowly with the magnetic field increase from $\sim$0.8 eV for $\sim 10^9$\,G (for the state $^2(-1)^+$) to $\sim$4.9 eV for $\sim 10^{11}$\,G\  (for the state $^4(-3)^+$). Hence, it may be visible in the infra-red/optical part of the spectra of radiation of a cold magnetic white dwarf.

\subsection{Acknowledgments}
This work was supported in part by PAPIIT grant {\bf IN115709} and CONACyT grant
{\bf 166189} (Mexico). The authors are deeply thankful to D. Turbiner (MIT - JPL NASA)
for designing the computer code prototype and for the creation of an optimal configuration
of a 96-processor cluster used for the calculations. The authors are also
obliged to E. Palacios for technical support.

\end{document}